\documentstyle[preprint,aps]{revtex}
\begin{document}
\title{Constraints on Coupling Constants through Charged $\Sigma$
Photoproduction}

\author{T.\ Mart}
\address{Institut f\"ur Kernphysik, Johannes Gutenberg-Universit\"at,\\
         55099 Mainz, Germany}

\author{C.\ Bennhold}
\address{Center of Nuclear Studies, Department of Physics,\\ The George
         Washington University, Washington, D. C. 20052}

\author{C. E.\ Hyde-Wright}
\address{Department of Physics, Old Dominion University,\\ Norfolk,
VA 23529-0116}

\date{\today}
\maketitle

\begin{abstract}
The few available data for the reactions $\gamma p \rightarrow
K^{0} \Sigma^{+}$ and $\gamma n \rightarrow K^{+} \Sigma^{-}$ are compared
to models developed for the processes $\gamma p \rightarrow K^{+}
\Sigma^{0}$ and $\gamma p \rightarrow K^{+} \Lambda$. It is found that
some of these phenomenological models overpredict the measurements by up to a
factor of 100. Fitting the data for all of these reactions leads to
drastically reduced Born coupling constants.
\end{abstract}
\pacs{13.60.Le, 25.20.Lj, 13.60.Rj}

Most analyses of kaon electromagnetic production over the last several
years have focused on the two processes $\gamma p \rightarrow K^{+}
\Lambda$ and $\gamma p \rightarrow K^{+}
\Sigma^{0}$\cite{abw,aw,tkb,benn,as1,cota}. This is clearly
due to the fact that almost all of the few available photokaon data have
been taken for these two reactions, along with the related kaon radiative
capture, $K^{-} p \rightarrow \Lambda \gamma$ and $K^{-} p \rightarrow
\Sigma^{0} \gamma$, and electroproduction processes. Despite the
considerable effort spent in the last years, kaon photoproduction on the
nucleon remains far from understood. Due to the limited set of data
the proliferating number of models permit only some qualitative conclusions
but do not yet allow the extraction of precise coupling constants and resonance
parameters. Meanwhile, however, a basic understanding of these elementary
reactions is required in order to predict cross sections for the
photoproduction of hypernuclei\cite{bw1}.

Most models to date are based on diagrammatic techniques using hadronic
degrees of freedom where a limited number of low-lying $s$-, $u$-, and
$t$-channel resonances are employed in a fit to the data along with the
standard set of Born terms. One general finding of all of these fits is
that the leading hadronic coupling constant $g_{K \Lambda N}$ cannot be
reconciled with the SU(3) value of
$3.0 < |g_{K \Lambda N}/\sqrt{4\pi}| < 4.4$
that is consistent with other hadronic
information such as $YN$ scattering. Instead, the value of $g_{K \Lambda N}$
extracted from photoproduction is too low unless a certain
$t$-channel resonance is included\cite{aw,as1} or absorptive factors are
applied
to reduce the Born terms\cite{tkb}. The latter method also helps to
eliminate the
divergence of these models at higher energies\cite{tkb}.  The SU(3)
range for the leading coupling constant in $\Sigma$ electromagnetic
production, $g_{K \Sigma N}$, is
$0.9 < |g_{K \Sigma N}/\sqrt{4\pi}| < 1.3$ which is compatible with the
range employed by the modern Nijmegen and Juelich YN-potentials. As
discussed in Refs.\cite{as1,work} the value of $g_{K \Sigma N}$
extracted from kaon photoproduction reactions varies widely and has
remained very uncertain.

In this note we develop extensions of previous models in order
to include the other four isospin channels listed in Table~\ref{channel}.
For this
purpose we employ the few available total cross section data for the charged
$\Sigma$-photoproduction reactions, $\gamma p \rightarrow K^{0} \Sigma^{+}$
and $\gamma n \rightarrow K^{+} \Sigma^{-}$.  The CEBAF
Large Acceptance Spectrometer (CLAS) will
detect neutral kaons and charged kaons with comparable efficiency
and will measure kaon photoproduction
on the neutron using deuterium\cite{schu,xiao}. Clearly
a theoretical study of the other isospin channels is called for.

In the electroproduction process for pseudoscalar mesons
 the transition matrix element can be written as:
\begin{eqnarray}
M_{fi} & = & \bar u(p_{Y}) \sum_{j=1}^6 A_{j}(k^{2},s,t) M_{j}~u(p_{N})
\end{eqnarray}
where $s$
and $t$ are the usual Mandelstam variables and $k^{2}$ denotes
the virtual photon momentum squared.
The gauge and Lorentz invariant matrices $M_{j}$ are given in many
references\cite{abw,as1,cota,deo,denn},
while the amplitudes $A_{j}$ can be obtained from Feynman diagrams.
For the photoproduction reactions only the amplitudes $A_{1} - A_{4}$
contribute.
For the vertex factors and propagators, we follow Ref.\cite{abw} with slight
modifications in order to ensure gauge invariance in the electroproduction
process.
We use pseudoscalar (PS), rather than pseudovector (PV) since previous
studies\cite{benn,deo,bw2,cohen} indicated the PS-coupling mode to be the
preferred one.  Chiral symmetry arguments that demand PV-coupling for
pions most likely do not apply to kaons due to their larger mass.

To relate the coupling constants in the Born terms among the various
isospin channels we first
consider the hadronic vertices. Since the Lambda is an
SU(3) isosinglet, one obtains
\begin{eqnarray}
g_{\Lambda} \equiv g_{K^{+} \Lambda p} = g_{K^{0} \Lambda n}
\end{eqnarray}
Similarly for the $N \longrightarrow K^{*} \Lambda$ vertex
\begin{eqnarray}
 g_{K^{*+} \Lambda p}^{V,T} = g_{K^{*0} \Lambda n}^{V,T}
\end{eqnarray}
On the other hand the Sigma is an isovector, so the $N \longrightarrow K
\Sigma$ couplings are related by the Clebsch-Gordan coefficients coupling
isospin 1 plus isospin 1/2 to isospin 1/2.
\begin{eqnarray}
 g_{\Sigma} \equiv g_{K^{+} \Sigma^{0} p} = -g_{K^{0} \Sigma^{0} n}
 = g_{K^{0} \Sigma^{+} p}/ \sqrt{2} = g_{K^{+} \Sigma^{-} n}/ \sqrt{2}
\end{eqnarray}

Note that there are different conventions used in expressing this
relation. Hadronic reactions commonly employ the definition of, e.g., Refs.
\cite{pil,dum}, where the isospin state of the $\Sigma^+$ is given as
$\Sigma^{+} = +|I=1,I_3=1>$.  In this paper, we define
$\Sigma^{+} = -|I=1,I_3=1>$ which is consistent with $Y_{l,m}^{*}=(-1)^m
Y_{l,-m}$ since $\Sigma^{-} = |I=1,I_3=-1>$. This convention
is customarily used in all meson photoproduction
reactions \cite{do}.

In $K^{0}$ photoproduction the vector meson exchanged in the t-channel
 is the $K^{*0} (896.1)$, hence the
transition moment $g_{K^{*}K \gamma}$ in $K^{+}$ production case
 must be replaced by the neutral
transition moment. The transition moment is related to the decay
width by \cite{thom}
\begin{eqnarray}
\Gamma_{K^{*} \rightarrow K \gamma} = \frac{9.8 MeV}{4 \pi} \vert
g_{K^{*} K \gamma} \vert^{2}
\end{eqnarray}
The measured decay widths are \cite{pdg}
\begin{eqnarray}
\Gamma_{K^{*+} \rightarrow K^{+} \gamma} = 50 \pm 5 keV
\end{eqnarray}
\begin{eqnarray}
\Gamma_{K^{*0} \rightarrow K^{0} \gamma} = 117 \pm 10 keV
\end{eqnarray}
Inserting these numbers in Eq.(5) we obtain
\begin{eqnarray}
\vert g_{K^{*0} K^{0} \gamma}/g_{K^{*+} K^{+} \gamma} \vert =
1.53 \pm 0.20
\end{eqnarray}
However, the sign of this ratio is undetermined experimentally. For the
phase of the neutral decay mode, we turn to a quark model
prediction, in particular,
the cloudy bag model computation by Singer and Miller \cite{sing}, which
accurately reproduces the experimental widths of Eqs.(6) and (7).
The quark and pion cloud
terms contribute in-phase to the $K^{*}$ photo decay, with the $K^{*0}$
amplitude of opposite sign as the $K^{*+}$ amplitude. Thus in our
photoproduction amplitudes we use
\begin{eqnarray}
g_{K^{*0} K^{0} \gamma} = -1.53 \hspace{2mm} g_{K^{*+} K^{+}
 \gamma}
\end{eqnarray}

Each of the nucleon resonances will be excited by an anomalous magnetic moment
\mbox{$\mu^{*} = (e/2M_{N}) \kappa^{*}$}, which can be written in terms of the
helicity amplitude $A_{1/2}$. The $N^{*} \rightarrow N \gamma$ decay width
can be used to relate the two quantities to each other,
\begin{eqnarray}
\Gamma_{N^{*} \rightarrow N \gamma} & = & \frac{\alpha}{4 \pi}
(\kappa^{*})^{2} \hspace{2mm} \frac{k_{c.m.}^{3}}{M^{2}}\nonumber\\
 & = & \frac{k_{c.m.}^{2}}{\pi} \hspace{2mm}
\frac{M}{M^{*}} \hspace{2mm} \vert A_{1/2} \vert^{2}
\end{eqnarray}
Thus
\begin{eqnarray}
\kappa_{n}^{*}/ \kappa_{p}^{*} = A_{1/2}^{n} / A_{1/2}^{p}
\end{eqnarray}
where we have used the quark model calculations by Koniuk
and Isgur \cite{kon} to constrain
the magnitude of the neutron amplitudes (see Table~\ref{magn}).
Thus, the coupling constants $G_{K \Lambda N^{*}} = g_{K \Lambda
N^{*}} g_{\gamma N N^{*}} $ for kaon production on the proton
are adjusted to the data, while for kaon production on the neutron
the couplings are multiplied with
the factor of Eq. (11) and the appropriate isospin factors.

The $\Sigma$ photoproduction reactions allow $\Delta$ resonance
contributions whose various coupling constants are
related by
\begin{eqnarray}
\hspace{10mm} G_{\Delta} \equiv G_{K^{+} \Sigma^{0} \Delta^{+}} =
-\sqrt{2} G_{K^{0} \Sigma^{+} \Delta^{+}} =
G_{K^{0} \Sigma^{0} \Delta^{0}} = \sqrt{2} G_{K^{+} \Sigma^{-} \Delta^{0}}
\end{eqnarray}
Here we used the same isospin convention as the one in Eq. (4).

As a first step, we limit our analysis to the four $\Sigma$ production
channels.  For the more qualitative findings presented here this proves
to be sufficient.  A more complete quantitative analysis that will
include the $\Lambda$ channels along with new upcoming data from Bonn
\cite{bonn} will be presented in a future work .
Here, we emphasize the fact that combining the
$K^+$$\Sigma^0$ photo- and electroproduction data \cite{data}
(86 and 96 points below 2.2 GeV, respectively) with the few total cross
section measurements of the
$K^{+} \Sigma^{-}$ and $K^{0} \Sigma^{+}$
channels in a common fit leads to very strong constraints on the
leading coupling constants.  To our knowledge, previous authors have not
included the charged $\Sigma$ channels in their analyses.

As in previous studies, our model includes the standard Born terms along
with the intermediate $\Lambda$- and $K^*$-exchange.  Furthermore, we
have incorporated the $N^*$ resonances $S_{11}$(1650) and $P_{11}$(1710) as
well as the $\Delta$ resonances $S_{31}$(1900) and $P_{31}$(1910) which can
only contribute to $\Sigma$ photoproduction. Our choice of resonance was
guided by our goal to draw qualitative conclusions about the behaviour
of coupling constants with a simple model that contains as few
parameters as needed to achieve a reasonable $\chi^2$. We found that
once a resonance with a particular spin-parity structure has been included
in the fit, adding additional states with the same quantum numbers
does not significantly reduce the $\chi^2$ any more. The
$S_{11}$(1650) and $P_{11}$(1710) states have considerable branching
ratios into the $K Y$ channels and, along with the
$S_{31}$(1900) and $P_{31}$(1910) $\Delta$ resonances, yielded the
smallest $\chi^2$ with a miminum number of parameters. We found that
adding additional resonances like the $S_{31}$(1620)
or the hyperonic $\Lambda^*$(1405) did not affect our conclusions.
Furthermore, our fit
does not contain the $K_1$(1270) t-channel resonance.  In contrast to
$K^{+} \Lambda$ production where the inclusion of this state led to a
K$\Lambda$N coupling constant in agreement with SU(3), we found no
sensitivity to this resonance
in K$\Sigma$ production.  In addition, due to the lack of
information on the $K_{1} \rightarrow K \gamma$ widths the $K_1$
contribution to the $K^0$ channel cannot easily be related to the $K^+$
channel.
With the current data base there clearly
remains an ambiguity as to which are the most
important resonances contributing to the kaon photoproduction process.
Future high precision data from CEBAF are expected to resolve this
issue.

Fig. 1 compares the predictions of three different models for the total
cross section of the four possible channels in K$\Sigma$ photoproduction.
The simplest model shown is taken from Ref. \cite{benn},
it contains only the Born
terms plus one additional $\Delta$-resonance at 1700 MeV and was fitted
solely to the $K^+$$\Sigma^0$ photoproduction data.
Furthermore, two of our new fits are shown, one includes the
$K^{+}\Sigma^{-}$ and $K^{0}\Sigma^{+}$ data, while the other one does not.
The coupling constants of the three models are given in Table \ref{cc}.

Fig. 1 clearly demonstrates that different models which give an
adequate description of the
$\gamma p \rightarrow K^{+} \Sigma^{0}$ data can give drastically
divergent predictions for the other isospin channels. The difference in the
Born coupling constants listed in Table III helps
to shed some light on these discrepancies. The model of Ref.\cite{benn}
that overpredicts the charged $\Sigma$ cross sections by up to two
orders of magnitude
contains the largest Born coupling constant.
The different predictions of our new model
with set II and set III of the coupling constants illustrate the same point.
Fitting all $K^{+} \Sigma^{0}$ photo- and electroproduction data (set II) leads
very large discrepancies with the $K^{+} \Sigma^{-}$ and
$K^{0} \Sigma^{+}$ total cross section data. Including those data into the
fit yields a coupling strength $g_{K \Sigma N}$ that differs by almost a
factor of 10 from the coupling constant in set II. Thus, fitting all data
simultaneously reduces the Born couplings to very small values, almost
eliminating the Born terms.  Clearly, the extracted couplings are
significantly below their SU(3) range as well as the values obtained in
hadronic reactions. This may be due to the neglect of hadronic form
factors at the strong interaction vertices, thus affecting especially
the non-resonant Born terms which are far off-shell even near threshold.
The SU(3) predictions, on the other hand, relate on-shell couplings
while determinations from low-energy hadronic scattering reactions
generally include form factors explicitly.  Future kaon photoproduction
studies will have to
address this question by including hadronic form factors in a gauge
invariant fashion.

We have compared a wide variety of models that are available in the
literature and always found the same pattern. For example, one of the
more advanced models developed in Ref. \cite{cota} was fitted to photo- and
electroproduction data of the $K^{+}\Lambda$, $K^{+}\Sigma^{0}$, and
$K^{+}\Lambda^{*}(1405)$ final states, while neglecting the charged
$\Sigma$ channels. Furthermore, they include crossing constraints
to simultaneously reproduce the $K^-$ radiative capture branching
ratios.  Their fit\cite{cota}
yielded significantly reduced couplings and its disagreement with the
experimental $K^{+} \Sigma^{-}$ and $K^{0} \Sigma^{+}$ data
is not as dramatic.

The underlying reason for the drastic differences in the various
predictions is elucidated in Fig. 2. Analyzing the individual diagramatic
contributions of the model of Ref.\cite{benn} in detail for the process
$\gamma p \rightarrow K^{+} \Sigma^{0}$, reveals
that the total cross section results from
successive destructive interferences between the various diagrams. The basic
Born terms consisting of the $K^{+}~ t$-channel, the $\Sigma^{0}~ u$-channel,
and the $p~ s$-channel exchanges governed by $g_{K \Sigma N}$ diverge very
quickly, adding the $\Lambda$ in the $u$-channel and the $K^{*}$ in the
$t$-channel leads to cancellations that reduce the calculated cross section
by up to an order of magnitude at higher energies. In contrast to the
$K^{+}\Sigma^{0}$ channel, the other three processes do not exhibit successive
destructive interferences,
leading to large predictions for the total cross sections. This
behavior can be traced to the relations between the coupling constants
in Eqs.(4) and (9). The
magnitude of the calculated charged $\Sigma$ cross sections is due in part to
the fact that the intermediate $u$-channel $\Lambda$ diagram cannot
contribute. Fitting all available data with one amplitude leads to the
observed drastic reduction
in the Born couplings, thus yielding a resonance dominated process.

In conclusion, we have shown that existing models for $K^+$$\Lambda$ and
$K^+$$\Sigma^0$ production can dramatically overpredict the few available
total cross section data for $K^{+} \Sigma^{-}$ and $K^{0} \Sigma^{+}$
photoproduction.  Including these data in the fit leads to drastically
reduced Born coupling constants $g_{K \Sigma N}$ and $g_{K \Lambda N}$,
yielding a description of the process that is resonance dominated.  It
is therefore imperative that future analyses include the complete data
base and that ongoing and upcoming experimental efforts provide data for
all isospin channels.

We are grateful to Prof. D. Drechsel and Dr. L. Tiator for helpful
discussions. TM is indebted
to Prof. D. Kusno, who suggested this work formerly. The work from TM is
supported by Deutscher Akademischer Austauschdienst and Deutsche
Forschungsgemeinschaft (SFB 201). CB is supported by the US DOE grant
no. DE-FG-05-86-ER40270 while CEH-W was supported by US DOE grant no.
DE-FG06-90ER40537.

\begin{table}
\caption{ The six reactions of photokaon production with their treshold
energies.}
\begin{tabular}{lcc}
 Type & $E_{\gamma,lab}^{tresh}$ (MeV) & $E_{tot.,c.m.}^{tresh}$ (MeV)\\
\tableline
 $ \gamma$ p $\rightarrow$  $K^{+} \Lambda$ & 911. & 1609\\
 $ \gamma$ n $\rightarrow$  $K^{0} \Lambda$ & 915.   & 1613\\
 $ \gamma$ p $\rightarrow$  $K^{+} \Sigma^{0}$ &  1046 & 1686\\
 $ \gamma$ p $\rightarrow$  $K^{0} \Sigma^{+}$ &  1048 & 1687\\
 $ \gamma$ n $\rightarrow$  $K^{+} \Sigma^{-}$ & 1052  & 1691\\
 $ \gamma$ n $\rightarrow$  $K^{0} \Sigma^{0}$ & 1051  & 1690\\
\end{tabular}
\label{channel}
\end{table}

\begin{table}
\caption{ $N + \gamma \rightarrow N^{*}(\frac{1}{2} \pm)$ Amplitudes}
\begin{tabular}{ccccccccc}
 Resonance & &   $J^{\pi}$ &&$A_{1/2}^{p}$(GeV$^{-1/2}$)& &$A_{1/2}^{n}$
(GeV$^{-1/2}$)& &
     $\kappa_{n}/ \kappa_{p}$  \\
 \tableline
$N(1650)$ && $\frac{1}{2}^{-}$ &&$ +88.$ &&$ -35.$ &&$ -0.40 $\\
$N(1710)$ && $\frac{1}{2}^{+}$ &&$ -47.$ &&$ -21.$ &&$ +0.45 $\\
[0.5ex]
\end{tabular}
\label{magn}
\end{table}

\begin{table}
\caption{Coupling constants (CC) set I comes from Ref. \protect\cite{benn},
         set II is generated by fitting to all  but  the charged
         $\Sigma$ data, and set III comes from fitting all available
         data in $K \Sigma$ production.}
\begin{tabular}{lccc}
 CC   & I  & II & III  \\
\tableline
$g_{K \Sigma N}/\sqrt{4\pi}$&$2.72 $ &1.30&0.130  \\
$g_{K \Lambda N}/\sqrt{4\pi}$&$-1.84$ &$-0.842$ &0.510\\
$G_{V}(K^{*})/4\pi$ &0.104&0.053 &0.052 \\
$G_{T}(K^{*})/4\pi$ &0.005&0.019 &0.053\\
$G_{N1}(1650)/\sqrt{4\pi}$ &-&$-0.136$  &0.111 \\
$G_{N2}(1710)/\sqrt{4\pi}$ &-&$-0.739$  &0.807\\
$G_{\Delta (1/2)}(1900)/\sqrt{4\pi}$ &-&0.125&0.109  \\
$G_{\Delta (1/2)}(1910)/\sqrt{4\pi}$ &-&0.746&0.457 \\
$G_{\Delta (3/2)}^{1}(1700)/4\pi$ &$-0.069$&-&- \\
$G_{\Delta (3/2)}^{2}(1700)/4\pi$ &0.314&-&- \\
$\chi^{2}/N$ &3.15&2.67 &5.30 \\
\end{tabular}
\label{cc}
\end{table}

\begin{figure}
\caption{Total cross section for the four isospin channels in $\Sigma$
         photoproduction. The dash-dotted curve represents the model
         with coupling constants of set I (Ref. \protect\cite{benn})
         in Table III.
         The dashed
         curve represents set II,
         while the solid curve shows the result from set III. For the
         $n(\gamma, K^{+})\Sigma^{-}$
         and $p(\gamma, K^{0})\Sigma^{+}$ graphs, the dash-dotted curve
         has been renormalized by a factor of 0.1 in order to fit on the
         scale.
         The experimental data are from \protect\cite{data}. }
\label{kps0}
\end{figure}

\begin{figure}
\caption{Contributions from the individual Born diagrams of the model from
         Ref.\protect\cite{benn} in the total $K \Sigma$ cross section.
         The dotted curve shows the basic $N$~$s$-channel,
         $\Sigma$~$u$-channel and $K$~$t$-channel diagrams only. The dashed
         curve includes the intermediate $\Lambda$ $u$-channel (only for
         $\Sigma^{0}$ production), the dash-dotted curve includes the
         $K^{*}$, while the solid curve shows the full model.
         For the $n(\gamma, K^{+})\Sigma^{-}$
         and $p(\gamma, K^{0})\Sigma^{+}$ graphs, all of the curves have
         been renormalized by a factor of 0.1 in order to fit on the
         scale.}
\label{k0sp}
\end{figure}

\begin{references}
\bibitem{abw}
R. A. Adelseck, C. Bennhold, and L. E. Wright, Phys. Rev {\bf C32}, 1681
(1985).
\bibitem{aw}
R. A. Adelseck and L. E. Wright, Phys. Rev. {\bf C38}, 1965 (1988).
\bibitem{tkb}
H. Tanabe, M. Kohno, and C. Bennhold, Phys. Rev. {\bf C39}, 741 (1989).
\bibitem{benn}
C. Bennhold, Phys. Rev. {\bf C39}, 1944 (1989).
\bibitem{as1}
R. A. Adelseck and B. Saghai, Phys. Rev. {\bf C42}, 108 (1990); Phys.
Rev. {\bf C45}, 2030 (1992).
\bibitem{cota}
R. A. Williams, C. R. Ji, and S. R. Cotanch, Phys. Rev. {\bf C46}, 1617
(1992).
\bibitem{bw1}
C. Bennhold and L.E. Wright, Phys. Rev. {\bf C39}, 927 (1989).
\bibitem{work}
R.L. Workman, Phys. Rev. {\bf C39}, 2456 (1989); Phys. Rev. {\bf44}, 552
(1991).
\bibitem{schu}
"Electromagnetic Production of Hyperons", CEBAF experiment E-89-004, R.
Schumacher (spokesman), and "Study of Kaon Photoproduction on
Deuterium", CEBAF experiment E-89-045, B. Mecking (spokesman)
\bibitem{xiao}
Xiaodong Li, L.E. Wright, and C. Bennhold, Phys. Rev. {\bf C45}, 2011
(1992).
\bibitem{deo}
B. B. Deo and A. K. Bisoi, Phys. Rev. {\bf D9}, 288 (1974).
\bibitem{denn}
P. Dennery, Phys. Rev. {\bf 124}, 2000 (1961).
\bibitem{bw2}
C. Bennhold and L.E. Wright, Phys. Rev. {\bf C36}, 438 (1987).
\bibitem{cohen}
J. Cohen, Int. J. Mod. Phys. {\bf A4}, 1 (1989).
\bibitem{pil}
H. Pilkuhn, {\it The Interactions of Hadrons} (North-Holland, Amsterdam,
1967).
\bibitem{dum}
O. Dumbrajs et al., Nucl. Phys. {\bf B216}, 277 (1983).
\bibitem{do}
A. Donnachie, Photo- and Electroproduction Processes in {\it High Energy
Physics}, Vol. 5, edited by E. H. S. Burhop (Academic Press, New York, 1972).
\bibitem{thom}
H. Thom, Phys. Rev. {\bf 151}, 1322 (1966).
\bibitem{pdg}
Particle Data Group, Phys. Rev. {\bf D45}, Part 2 (1992).
\bibitem{sing}
P. Singer and G. A. Miller, Phys. Rev. {\bf D33}, 141 (1986).
\bibitem{kon}
R. Koniuk and N. Isgur, Phys. Rev. {\bf D21}, 1868 (1980).
\bibitem{bonn}
M. Bockhorst et al., Z. Phys. {\bf C63}, 37 (1994).
\bibitem{data}
Photoproduction of Elementary Particles, Vol. 8 of {\it Landolt-B\"ornstein
Numerical Data and Functional Relationships in Science and Technology},
edited by H. Genzel, P. Joos, and W. Pfeil (Springer, New York 1973),
p. 295-298, 336.\\
P. Brauel et al., Z. Phys. {\bf C3}, 101 (1979) and references therein.
\end{references}
\end{document}